\documentclass[prx,aps,twocolumn,preprintnumbers,superscriptaddress,longbibliography,nofootinbib, 10pt]{revtex4-2}

\usepackage{graphicx,subeqnarray}

\usepackage{url}
\usepackage{bm}
\usepackage{tikz} 
\usepackage{pifont} 
\usepackage{amsfonts}
\usepackage{algorithmicx}
\usepackage[noend]{algpseudocode}
\usepackage[most]{tcolorbox}

\algrenewcommand\algorithmicrequire{\textbf{Input:}}
\algrenewcommand\algorithmicensure{\textbf{Output:}}

\newcounter{algbox}
\newenvironment{AlgBox}[1]{%
  \refstepcounter{algbox}%
  \begin{tcolorbox}[enhanced,breakable,sharp corners,boxrule=.4pt,colback=white,
    title=\textbf{Algorithm \thealgbox}\ #1]}{\end{tcolorbox}}
\usepackage{amssymb}
\usepackage{amsmath,upgreek}
\usepackage{blindtext}
\definecolor{xlinkcolor}{cmyk}{1,0.6,0,0}
\newcounter{appctr}
\renewcommand{\theappctr}{\Alph{appctr}}
\newcommand{\appsection}[2]{%
  \refstepcounter{appctr}%
  \section*{\theappctr. #1}%
  \label{#2}%
}

\usetikzlibrary{positioning, shapes.geometric, arrows.meta, calc,fit,backgrounds}

\usepackage{silence}
\usepackage[bookmarks=false,         
     pdfnewwindow=true,      
     colorlinks=false,    
     linkcolor=xlinkcolor,     
     citecolor=xlinkcolor,     
     filecolor=xlinkcolor,  
     urlcolor=xlinkcolor,      
final=true]{hyperref}

\usepackage{subcaption}
\begin{document}


\title{BOA Constrictor: A Mamba-based lossless compressor for High Energy Physics data}


\author{Akshat Gupta}
\email{akshat.gupta-4@postgrad.manchester.ac.uk}
\affiliation{Department of Physics \& Astronomy, University of Manchester, Manchester M13 9PL, United Kingdom}
\affiliation{Centre for Quantum Science and Engineering, University of Manchester, Manchester M13 9PL, United Kingdom}

\author{Caterina Doglioni}
\email{caterina.doglioni@manchester.ac.uk}
\affiliation{Department of Physics \& Astronomy, University of Manchester, Manchester M13 9PL, United Kingdom}
\author{Thomas J.~Elliott}
\email{physics@tjelliott.net}
\affiliation{Department of Physics \& Astronomy, University of Manchester, Manchester M13 9PL, United Kingdom}
\affiliation{Centre for Quantum Science and Engineering, University of Manchester, Manchester M13 9PL, United Kingdom}
\affiliation{Department of Mathematics, University of Manchester, Manchester M13 9PL, United Kingdom}

\date{\today}
\begin{abstract}
The petabyte-scale data generated annually by High Energy Physics (HEP) experiments like those at the Large Hadron Collider present a significant data storage challenge. Whilst traditional algorithms like LZMA and ZLIB are widely used, they often fail to exploit the deep structure inherent in scientific data. We investigate the application of modern state space models (SSMs) to this problem, which have shown promise for capturing long-range dependencies in sequences. We present the Bytewise Online Autoregressive (BOA) Constrictor, a novel, streaming-capable lossless compressor built upon the Mamba architecture. BOA combines an autoregressive Mamba model for next-byte prediction with a parallelised streaming range coder. We evaluate our method on three distinct structured datasets in HEP, demonstrating state-of-the-art compression ratios, improving upon LZMA-9 across all datasets. These improvements range from 2.21$\times$ (vs. 1.69$\times$) on the ATLAS dataset to a substantial 44.14$\times$ (vs. 27.14$\times$) on the highly-structured CMS dataset, with a modest $\sim 4.5$MB model size. However, this gain in compression ratio comes with a trade-off in throughput; the Storage-Saving Rate ($\sigma_{SSR}$) of our prototype currently lags behind highly-optimised CPU-based algorithms like ZLIB. We conclude that while this Mamba-based approach is a highly promising proof-of-principle, significant future work on performance optimisation and hardware portability is required to develop it into a production-ready tool for the HEP community.
\end{abstract}

\maketitle 


\section*{INTRODUCTION}
As global demand for data storage and sharing continue to grow, managing such volumes has become increasingly challenging. This issue is particularly acute in high-energy physics (HEP), where vast and complex datasets routinely push the limits of existing compression and storage technologies: each year, experiments at the Large Hadron Collider (LHC) at CERN produce approximately thirty petabytes of data \cite{CERN_LHC_Facts}. 

Current solutions, such as the ROOT framework combined with algorithms like Lempel–Ziv–Markov chain Algorithm (LZMA) and ZLIB, are currently used to address these challenges \cite{philippe_canal_2025_16944732, Shadura_2020}. In this work, we investigate whether greater gains in storage efficiency can be achieved within the constraints of existing infrastructure with a focus on minimal data deterioration and loss \cite{10.1145/2503210.2503283}, through improved lossless compression methods.

Algorithms commonly employed for HEP data compression, such as LZMA and ZLIB, largely overlook the inherent structure of the data, including correlations between sequential measurements and dependencies among physical parameters \cite{Shadura_2020}. A more ``data-aware" compression can be achieved via the use of neural networks that learn the underlying patterns of data \cite{li2025losslessdatacompressionlarge, mao2025losslesscompressionlargelanguage, ratsaby2010predictioncompression,bamler2022understandingentropycodingasymmetric}. While this is not a new approach, early neural compressors based on Recurrent Neural Networks (RNNs) were computationally prohibitive. Later models based on the Transformer architecture showed very good performance but 
suffer from a quadratic complexity bottleneck ($O(L^2)$) \cite{vaswani2023attentionneed, toderici2017resolutionimagecompressionrecurrent, yang2023introductionneuraldatacompression}. 
The recent development of the Mamba State Space Model (SSM) \cite{dao2024transformersssmsgeneralizedmodels, gu2024mambalineartimesequencemodeling} has fundamentally changed this landscape. 
Mamba combines the strengths of RNNs and Transformers, offering a model that can process sequences with linear time complexity and constant memory during inference \cite{dao2024transformersssmsgeneralizedmodels, gu2024mambalineartimesequencemodeling}. They can also capture extremely long-range dependencies \cite{dao2024transformersssmsgeneralizedmodels, gu2024mambalineartimesequencemodeling}. This unique combination of features makes them a prime-candidate for a high-performance learned compressor.

This work presents the design, implementation, and evaluation of the Bytewise Online Autoregressive (BOA) Constrictor, a streaming lossless compressor built on the Mamba architecture. A modest $\sim 4.5$\,MB BOA model coupled to a parallelised range coder consistently outperforms LZMA-9 on three representative HEP datasets: from 1.69$\times$ to 2.21$\times$ on ATLAS and from 27.14$\times$ to 44.14$\times$ on CMS. This improved compression comes at the expense of lower throughput on current hardware, highlighting the deployment trade-offs for neural compressors in HEP.


\section*{Theoretical Framework}
\subsection*{Entropy-Based Compression and Cross-Entropy Loss}
Consider data sourced from an independent and identically-distributed (i.i.d.) distribution $P$, taking possible values $x\in\mathcal{X}$. The entropy $H(P)=-\sum_x P(x)\log_{256} P(x)$ \cite{6773024} quantifies the fundamental limit of lossless compression, in bytes, in the asymptotic limit. By the source–coding theorem for stationary ergodic sources, the expected code length per symbol satisfies $\tfrac{1}{n}\mathbb{E}[\ell(X_{1:n})]\ge H(P)$ bytes~\cite{6773024}, with equality achieved as $n\to\infty$ under optimal coding that uses $P$ \cite{b3dc75af-1a78-39c9-b837-c7f3ee48d5d5}. For a specific sequence $x$, the prefix Kolmogorov complexity $K(x)=\min_p\{|p|:U(p)=x\}$ implies that any lossless code $C$ must satisfy $|C(x)|\ge K(x)-O(1)$, whilst $\mathbb{E}[K(X_{1:n})]=nH(P)+O(1)$ \cite{b3dc75af-1a78-39c9-b837-c7f3ee48d5d5}.

When we approximate the true source distribution $P$ with a model distribution $Q$, the achievable compression is governed by the cross-entropy $H(P,Q)=H(P)+D_{\mathrm{KL}}(P\Vert Q)$, where $D_{\mathrm{KL}}(P\Vert Q)=\sum_xP(x)\log_{256}(P(x)/Q(x))$ is the Kullback-Leibler Distance \cite{Kullback1951OnIA}. The optimal expected code length per symbol under $Q$ is $H(P,Q)$ bytes, reaching the minimum $H(P)$ iff $Q=P$ \cite{Kullback1951OnIA}. This information-theoretic perspective provides the foundation for our compression approach. 

We serialise each table row into a causal byte sequence $\mathbf{y}=(y_1,\ldots,y_L)\in\{0,\ldots,255\}^L$ using a decoder-safe grammar $G$. An entropy-based compressor comprises two components: a predictor $Q$ that assigns conditional probabilities $Q(y_t=k \mid y_{<t})$ over $k\in\{0,\ldots,255\}$, conditioned only on the preceding bytes; and an entropy coder $C$ that transforms the pair $(\{q_t\}_{t=1}^{L}, \mathbf{y})$ into a compressed bytestream, where $q_t := Q(y_t \mid y_{<t})$ is the probability assigned to the realised byte.

The ideal compression length (in bytes) under the true distribution $P$ would be $-\log_{256} P(y_t \mid y_{<t})$ at each position. In practice, using the predictor $Q$, the actual compression length is
\begin{equation}
\begin{aligned}
\text{Total Bytes} \;=\; \sum_{t=1}^{L} -\log_{256} Q(y_t \mid y_{<t}).
\end{aligned}
\end{equation}
That is, the average number of bytes per symbol is $H(P,Q)$, corresponding to the theoretical ideal $H(P)$ iff $Q=P$.

Traditional compressors approximate these probabilities using heuristics. Dictionary-based methods (e.g. LZMA \cite{1055714, 1055934}) implicitly assign high probability to sequences that have appeared previously in a sliding window, proving very effective at capturing literal and local redundancies. Statistical methods (e.g. Prediction by Partial Matching (PPM) \cite{61469, 1096090}) explicitly build a statistical model of symbol contexts to predict the next symbol.

In our approach, we employ a neural network trained as an autoregressive model to serve as the probability estimator $Q$. This work leverages the Mamba architecture as a state-of-the-art predictive model for this purpose. For byte tokens $y_t\in\{0,\ldots,255\}$ with model probabilities $p_i^{(t)}=Q(Y_t=i\mid Y_{<t})$, the per-step loss is
\begin{equation}
\begin{aligned}
    \mathcal{L}_{\mathrm{CE},t}=-\log_{256} Q(y_t\mid y_{<t}).
\end{aligned}
\end{equation}
The total sequence loss is $\mathcal{L}_{\mathrm{CE}}=\sum_{t=1}^{L}\mathcal{L}_{\mathrm{CE},t}$. Taking the expectation over the true source $P$, we obtain
\begin{equation}
\begin{aligned}
\mathbb{E}_{P}[\mathcal{L}_{\mathrm{CE}}=H(P,Q)]\\=H(P)+D_{\mathrm{KL}}(P\!\parallel\!Q)\ge H(P),
\end{aligned}
\end{equation}
with equality iff $Q=P$. Minimising this cross-entropy loss therefore directly optimises our model for compression, as reducing $D_{\mathrm{KL}}(P\Vert Q)$ simultaneously improves both predictive accuracy and compression efficiency.

\subsection*{Range coding}
To implement the entropy-based compression, our framework pairs an autoregressive neural predictor ($Q$) with an entropy coder ($C$). For the coder, we use range coding~\cite{Martin1979RangeEA}, which is an integer form of arithmetic coding that emits bytes as soon as they are fixed. It keeps an integer base $L$ and width $R$ so the active interval is $[L,\,L{+}R)$ on a $w$-bit ring. Symbols are encoded using an integer Cumulative Distribution Function (CDF).

For the predictor, we implement $Q$ using the Mamba state-space model (SSM)~\cite{gu2024mambalineartimesequencemodeling}, chosen for its ability to model long-range dependencies with linear-time complexity and constant memory during inference (see Appendix~\ref{mambabg} for a detailed background).

\paragraph*{Alphabet, sequence, and model.}
Let $\mathcal{A}=\{0,\dots,255\}$ and $s_{1:N}\in\mathcal{A}^N$. The autoregressive Mamba network exposes a \texttt{step} function that, given the previous byte $s_{t-1}$ and the current stream state $S_t$, returns logits:
\begin{equation}
\ell_t(a\mid s_{<t};\theta)\in\mathbb{R}^{256},\quad a\in\mathcal{A}.
\end{equation}
This model is implemented as an embedding layer, $L$ stacked Mamba blocks with a stream cache, and a final linear head. The state transition is deterministic:
\begin{equation}
(S_{t+1},\,\ell_{t+1})=\Phi_\theta(S_t,\,s_t).
\end{equation}

\paragraph*{Probabilities and integer CDF.}
The logits are converted to probabilities via softmax:
\begin{equation}
p_t(a\mid s_{<t})=\frac{\exp(\ell_t(a))}{\sum_{x\in\mathcal{A}}\exp(\ell_t(x))}.
\end{equation}
We choose a precision $B$ with $M=2^B$ and quantise the probabilities into a strictly positive histogram $f_t(a)\in\mathbb{N}$ such that $\sum_a f_t(a)=M$. From this, we define the integer CDF:
\begin{align}
C_t(a)&=\sum_{x<a} f_t(x),
\end{align}

\paragraph*{Encoder.}
Given the true byte $s_t$, the range coder updates its state $(L, R)$:
\begin{align}
L &\leftarrow L + \Big\lfloor \tfrac{R\,C_t(s_t)}{M} \Big\rfloor,\\
R &\leftarrow \Big\lfloor \tfrac{R\,\big(C_t(s_t{+}1)-C_t(s_t)\big)}{M} \Big\rfloor.
\end{align}
The coder then renormalises while the top byte is fixed. With $E=L+R-1$, if
\begin{equation}
\Big\lfloor \tfrac{L}{2^{w-8}}\Big\rfloor=\Big\lfloor \tfrac{E}{2^{w-8}}\Big\rfloor,
\end{equation}
it outputs that byte and rescales the interval:
\begin{equation}
L \leftarrow (L \ll 8)\bmod 2^w,\qquad R \leftarrow (R \ll 8)\bmod 2^w.
\end{equation}
Finally, the model state is advanced using the true $s_t$.

\paragraph*{Decoder.}
The decoder maintains the same $(L,R)$ state and the current code value $C\in[L,L{+}R)$. It first finds the target value $x$ within the scaled range:
\begin{equation}
x=\Big\lfloor \tfrac{(C-L+1)M-1}{R}\Big\rfloor\in\{0,\dots,M-1\},
\end{equation}
and then decodes the byte $s_t$ by finding the unique symbol such that $C_t(s_t)\le x < C_t(s_t{+}1)$. It applies the same $(L,R)$ update as the encoder. On each byte $b$ read from the compressed stream (which occurs during renormalisation), it updates the code value:
\begin{equation}
C \leftarrow \big((C \ll 8)+b\big)\bmod 2^w.
\end{equation}
Finally, the decoder advances its model state using the decoded $s_t$.

\paragraph*{Determinism and synchronisation.}
Using evaluation mode and the streaming cache ensures that both the encoder and decoder compute identical logits $\ell_t$, and thus identical integer CDFs $f_t$ and $C_t$. As the histogram bins are strictly positive and the normalisation $M$ is identical, both coder states remain in perfect lockstep.
\section*{Methodology}
\subsection*{Dataset}
We use three data files from the CERN Open Data portal \cite{cernopendata} to evaluate the performance of BOA, either as-is or pre-processed: (a) a data file from a CMS open dataset~\cite{CMS_collaboration2017-rb}, (b) a file from an ATLAS open dataset ~\cite{ATLAS_collaboration2025-vq}, (c) a  human-readable ASCII plain text in HEPMC format  \cite{Buckley_2021, ATLAS_collaboration2025-fq}, and (d) a smaller, skimmed and slimmed version of the CMS data file in numpy format \cite{CMS_collaboration2017-rb}. It is important to note that these are structured data formats with defined columns.

From the CMS data file we form a fixed-width subset of 50\,000 events with 5\,359 columns by zero padding and storing as an uncompressed \verb|bin| file ($2044.30$MB). For the ATLAS dataset, we form a subset of 5\,619\,475 jets with 30 columns ($546.63$MB). The HEPMC file is uncompressed and unaltered (1992.24MB). The last dataset is another slimmed CMS dataset which contains 24 jet features and 520\,000 events converted to \verb+float32+ and the column names discarded ($47.61$MB). This last dataset is provided in the code repository as a bundled reference. All subsequent analyses use these subsets to ensure exact reproducibility (see Appendix \ref{whybaler}). Baseline LZMA-9 compression ratios are 27.14$\times$ for CMS\footnote{The CMS dataset is already highly structured and repetitive at the byte level, which leads to a high baseline compression ratio with LZMA.}, 1.69$\times$ for ATLAS, 5.39$\times$ for HEPMC, and 3.22$\times$ for the Bundled CMS using LZMA library in Python \cite{pub:5007, 1055714, 1055934}. All datasets have been released as open data under the Creative Commons CC0 waiver by CERN. Events in these datasets, where applicable, are independent with no temporal correlation. For model training we take a further 200\,MB sample from the subset and split it into training, validation, and test sets in an 80\%/10\%/10\% ratio. This setup provides evidence of online compression of unseen but similar data.
\subsection*{Models}
\begin{figure}
    \centering
    \includegraphics[width=\linewidth]{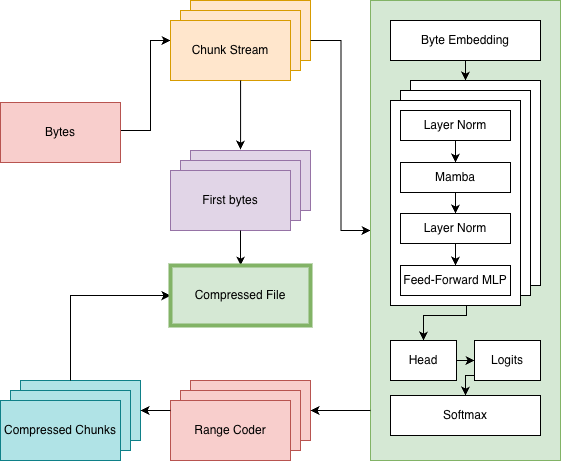}
    \caption{Conceptual diagram of our approach to the compression technique. Multiple boxes signify parallel streams except within the Model where they represent multiple blocks.}
    \label{fig:conceptimg}
\end{figure}
\begin{AlgBox}{Forward Pass}\label{alg:one}
\begin{algorithmic}[1]
  \Require $x$ (input seq), $d_m$ (model dim), $n_\ell$ (layers), $ip$ (inference params)
  \State $x \gets \textsc{ByteEmbed}(x,d_m)$
  \For{$\ell=1$ \textbf{to} $n_\ell$}
    \State $x \gets \textsc{MambaBlock}_\ell(x; ip)$
    \State $x \gets \textsc{LayerNorm}(\textsc{FFN}(x))$
  \EndFor
  \State $\text{logits} \gets \textsc{Head}(x)$
  \Ensure logits (next-byte predictions)
\end{algorithmic}
\end{AlgBox}
We introduce the \textit{BOABytePredictor} model which is designed to predict the next byte in a sequence. It uses an embedding layer for the 256 possible byte values, followed by $n$ Mamba layers \cite{mamba, mambapy}. The model output is a prediction of the next byte, with a final linear layer that maps the hidden representation to logits for each of the 256 possible next bytes.

The model consists of the following components:
\begin{enumerate}
    \item \textbf{Embedding Layer}: Converts input data into dense representations.
    \item \textbf{MambaBlock}: A key building block that consists of layer normalization, Mamba processing, and a feedforward network.
    \item \textbf{Head Layer}: The final output layer that produces logits for each possible value.
\end{enumerate}
The forward pass for the models is shown in Algorithm \ref{alg:one}. This model has two primary tunable hyper-parameters: i) the model dimension ($d_{m}$; and ii) the number of layers ($n_{l}$). These are fixed at $d_m = 256$ and $n_l = 1$ for an optimal balance between model size and compression ratios ($\sim 4.5$MB model size). Other tunable hyper-parameters include sequence length (default: 10000), batch size (default: 5), and learning rate (default: $0.0005$).

\subsection*{Coders \& Parallelisation}
We use the range coder from the \texttt{constriction} library \cite{bamler2022constriction} to encode the data based on the probabilities generated by the predictor models for CPU-based and a custom serial GPU range coder for GPU-based compression. This approach is ideal for our setup because it employs a stacking layout, where the first byte encoded is also the first byte decoded.

Autoregressive range coding is inherently sequential at the symbol level \cite{bamler2022constriction}, yet we obtain substantial speed-ups by parallelising across many independent streams and leveraging vectorised GPU kernels at a constant overhead (due to storing multiple first bytes).
As illustrated in Figure~\ref{fig:conceptimg}, during compression we split the input bytes into \(n\) chunks (streams).
Using a streaming state in the Mamba layer, we compute next-symbol probabilities per timestep for all active streams on the GPU and encode them in parallel with a GPU range coder.
We record in a \texttt{.boa} container the first byte of each stream, the compressed streams, the nominal chunk length \(\ell\), the length of the final chunk, the total uncompressed length, a compact model fingerprint, and a CRC-protected index of per-stream offsets and sizes.

During decompression we load this metadata, reconstruct per-timestep probabilities with the same model, and decode all streams in parallel on the GPU; streams that have finished are masked out at each step.

Both compression and decompression run under \texttt{torch.inference\_mode()} in the PyTorch package \cite{paszke2019pytorchimperativestylehighperformance} to eliminate autograd overhead.
On the CPU hot path\footnote{i.e., the sequence of instructions within a program that is executed exceptionally frequently.} we process different streams on separate threads using the same per-timestep procedure, which is slower than the GPU path as expected.
\section*{Streaming}
\begin{figure}[ht!]
  \centering
  \includegraphics[width=0.8\linewidth]{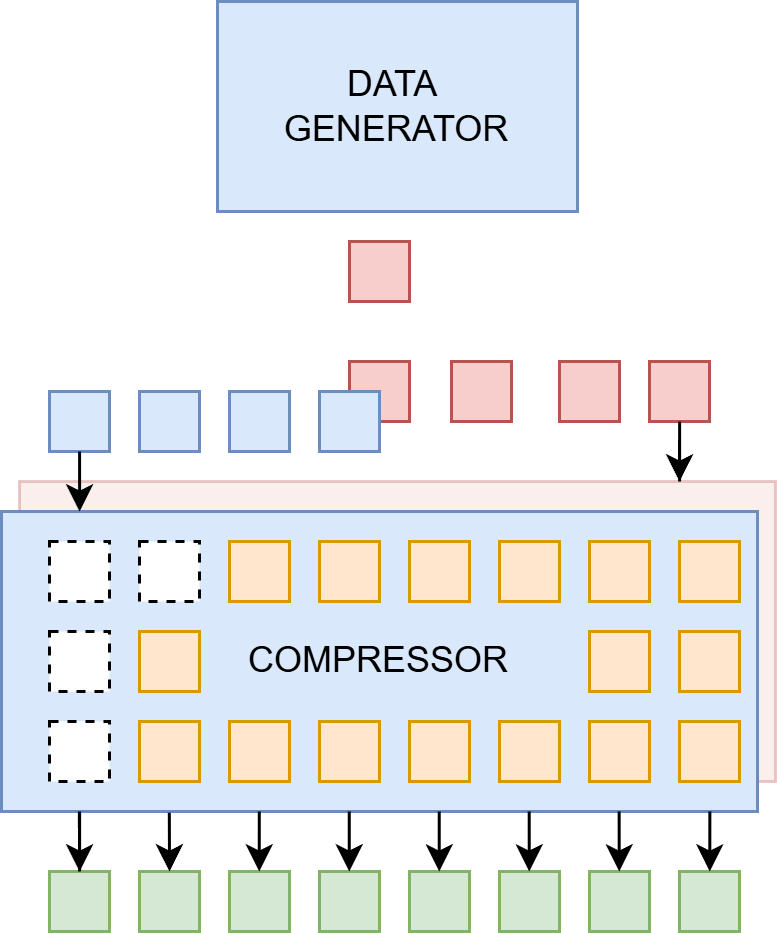}
  \caption{Streaming pipeline: buffer chunks in parallel (blue), compress in parallel (green), and pre-load the next compressor set (red).}
  \label{fig:concept2}
\end{figure}
Figure~\ref{fig:concept2} shows the streaming pipeline. The data generator emits a sequence that a distributor splits into fixed-size chunks buffered in parallel (blue). Once a lane fills, its chunk enters a worker that computes model probabilities and performs range coding to produce a shard (green). While active workers compress, the system preloads the next set of chunks into an idle group to hide latency and keep throughput high (red). Arrows indicate top-to-bottom flow; parallel lanes indicate concurrency. We demonstrate this through a pseudo-streaming prototype: we replay a pre-selected dataset, split it into fixed-size chunks, and run them through the BOA kernels.
\subsection*{Compression Metrics}

Let $N_{\mathrm{orig}}$ be the original file size, $N_{\mathrm{comp}}$ be the compressed file size, and $T$ be the compression time. 

Given that the algorithm provides lossless compression, two key performance metrics are:
\begin{itemize}
    \item \textbf{Compression Ratio (CR):} $CR = N_{\mathrm{orig}} / N_{\mathrm{comp}}$
    \item \textbf{Storage-Saving Rate ($\sigma_{SSR}$):} $\sigma_{SSR} = (N_{\mathrm{orig}}-N_{\mathrm{comp}}) / T$
\end{itemize}
\section*{Results \& Discussion} 
\begin{table*}[t]
    \centering
    \begin{tabular}{l|c|c|c|c}
        \hline \hline
        \textbf{Compressor} & \textbf{Size (MB)} & \textbf{Compression Ratio} & \textbf{Compression Time (s)} & \textbf{$\mathbf{\sigma_{SSR}}$}\\
        \hline \hline
        \textbf{CMS}\\
        \hline
        Baseline & 2044.30 & N/A & N/A & N/A\\
        LZMA(9) & 75.32 & 27.14 & 908.47 & 2.17\\
        ZLIB(9) & 116.11 & 17.61 &  \textbf{138.20} & \textbf{14.67} \\
        BOA & \textbf{46.32} & \textbf{44.14} & 514.25 & 3.89\\
        \hline
        \textbf{ATLAS}\\
        \hline
        Baseline & 546.63 & N/A & N/A & N/A\\
        LZMA(9) & 322.75 & 1.69 & 356.36 & 0.63\\
        ZLIB(9) & 338.86 & 1.61 &  \textbf{28.58} & \textbf{7.27} \\
        BOA & \textbf{247.33} &  \textbf{2.21} & 140.41 & 2.31\\
        \hline
        
        \textbf{HEPMC}\\
        \hline
        Baseline & 1992.24 & N/A & N/A & N/A\\
        LZMA(9) & 369.69 & 5.39 & 2425.31 & 0.67 \\
        ZLIB(9) & 497.83 & 4.00 & \textbf{156.75} & \textbf{9.53} \\
        BOA & \textbf{221.59} &  \textbf{8.99} & 506.78 & 3.50\\
        \hline
        \textbf{Bundled CMS}\\
        \hline
        Baseline & 47.61 & N/A & N/A & N/A\\
        LZMA(9) & 14.73 & 3.22 & 25.96 & 1.27\\
        ZLIB(9) & 18.57 & 2.56 &  \textbf{10.29} & 2.82\\
        BOA & \textbf{11.81} &  \textbf{4.03} & 12.26 & \textbf{2.92}\\
        \hline
    \end{tabular}
    \caption{Comparison of popular compression algorithms against Boa Constrictor (BOA). The compression times exclude training time.}
    \label{tab:comparison}
\end{table*}

In Table~\ref{tab:comparison} we compare ROOT-standard compressors with BOA. All BOA models train for 10 epochs at about \(70\,\mathrm{s}\) per epoch with fixed bytes and split ratios, except the smaller bundled CMS which uses \(80\%\) of the data and trains at about \(26\,\mathrm{s}\) per epoch. Compression runs with \(10{,}000\) chunks and \(5{,}000\) concurrent streams to demonstrate streaming and fit GPU memory. Training time is excluded from table timings.

BOA achieves the best compression ratio on all four datasets. However, referencing the storage-saving rate \(\sigma_{\mathrm{SSR}}\), ZLIB-9 outperforms BOA except on the bundled CMS. The gap reflects symbol-sequential coding with GPU matrix multiplications, compared to hand-tuned CPU vector code. Larger chunks, tuned kernels, and more GPUs should raise the \(\sigma_{\mathrm{SSR}}\) by amortising per-chunk overheads across lanes.
\begin{figure*}[ht!]
  \centering
  
  \begin{subfigure}{0.49\linewidth}
    \centering
    \includegraphics[width=0.8\linewidth]{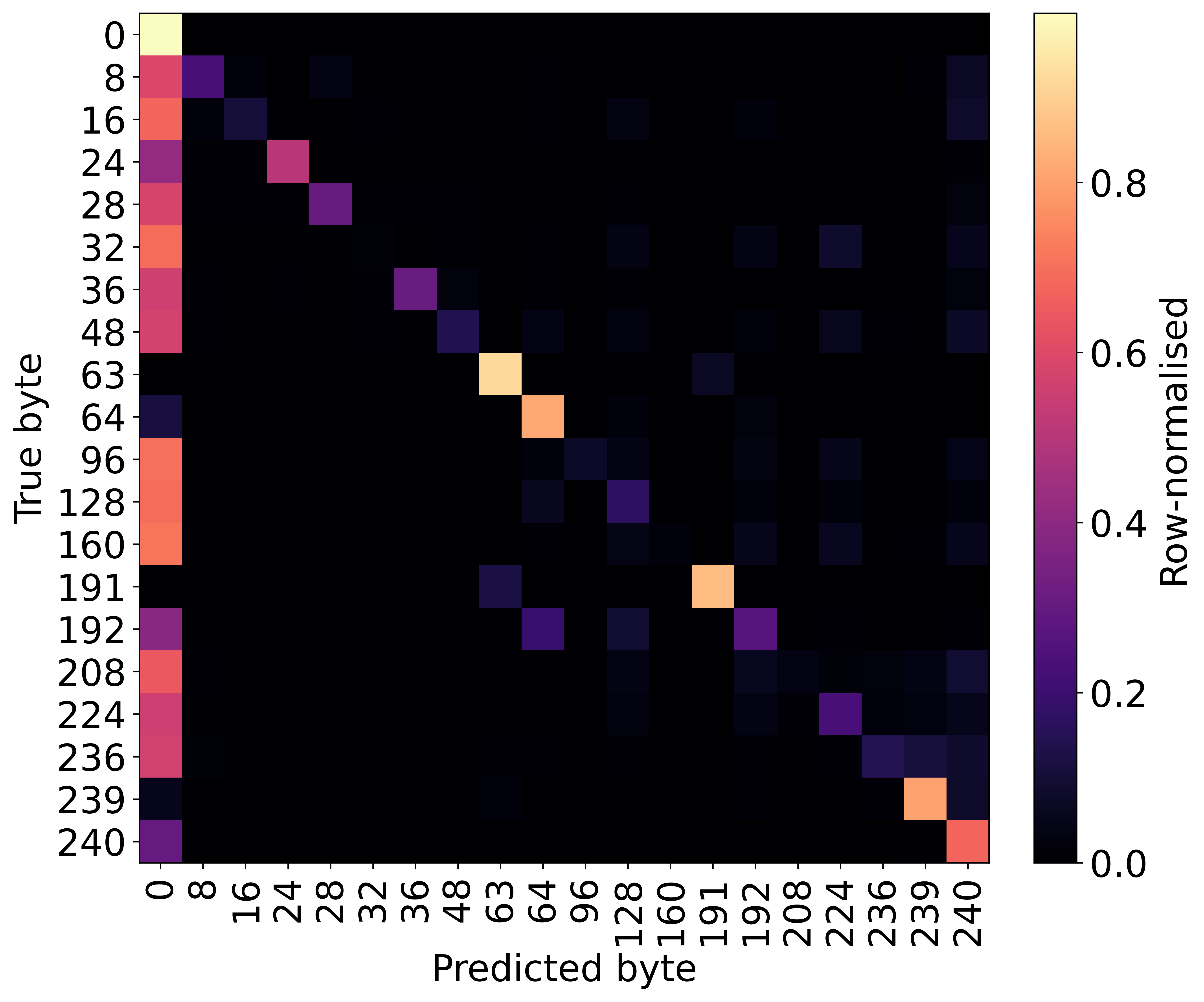}
    \caption{CMS}
    \label{fig:ConfMatrix:a}
  \end{subfigure}
    \begin{subfigure}{0.49\linewidth}
    \centering
    \includegraphics[width=0.8\linewidth]{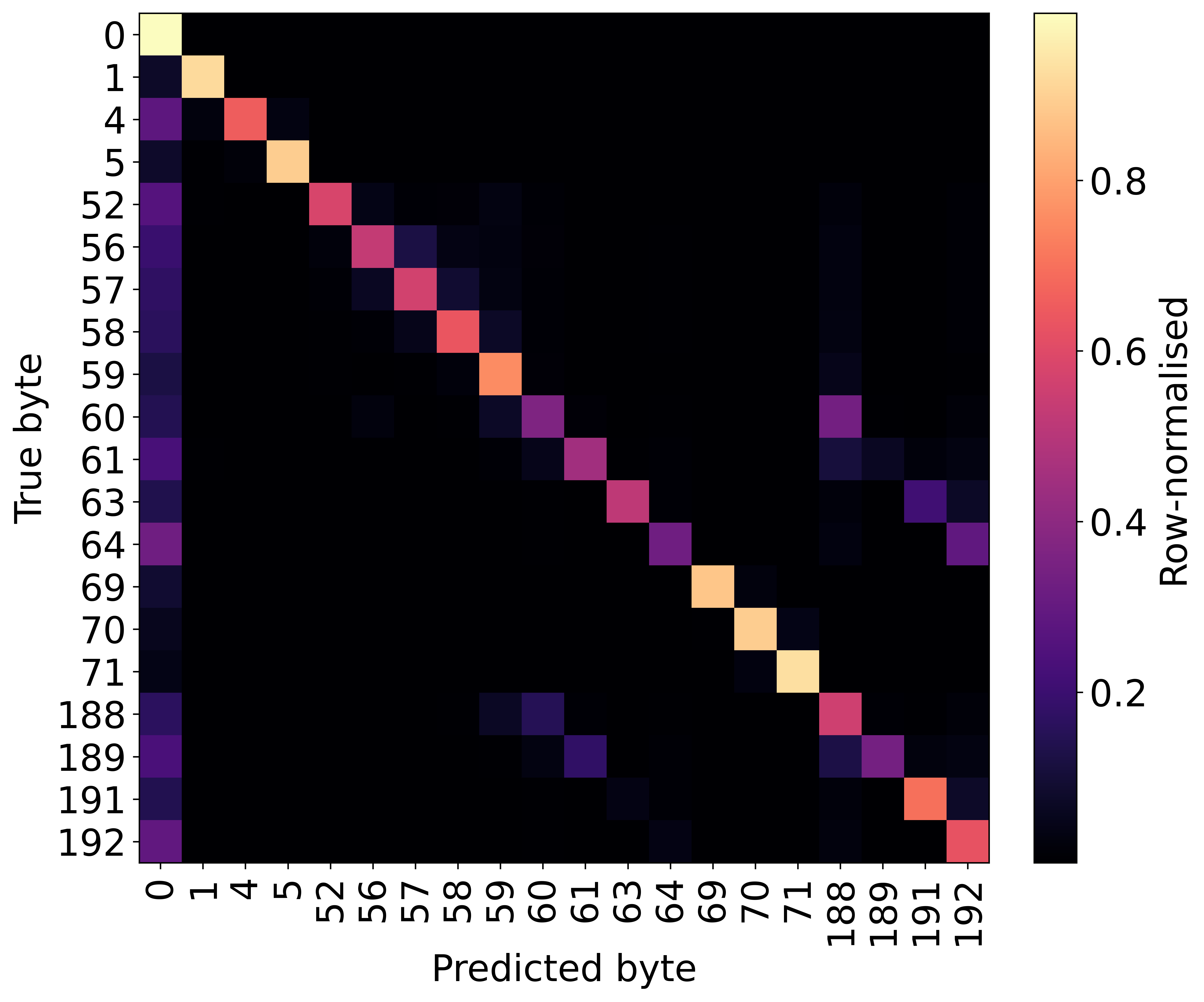}
    \caption{ATLAS}
    \label{fig:ConfMatrix:b}
    \end{subfigure}
  \vspace{0.6em}

  \begin{subfigure}{0.49\linewidth}
    \centering
    \includegraphics[width=0.8\linewidth]{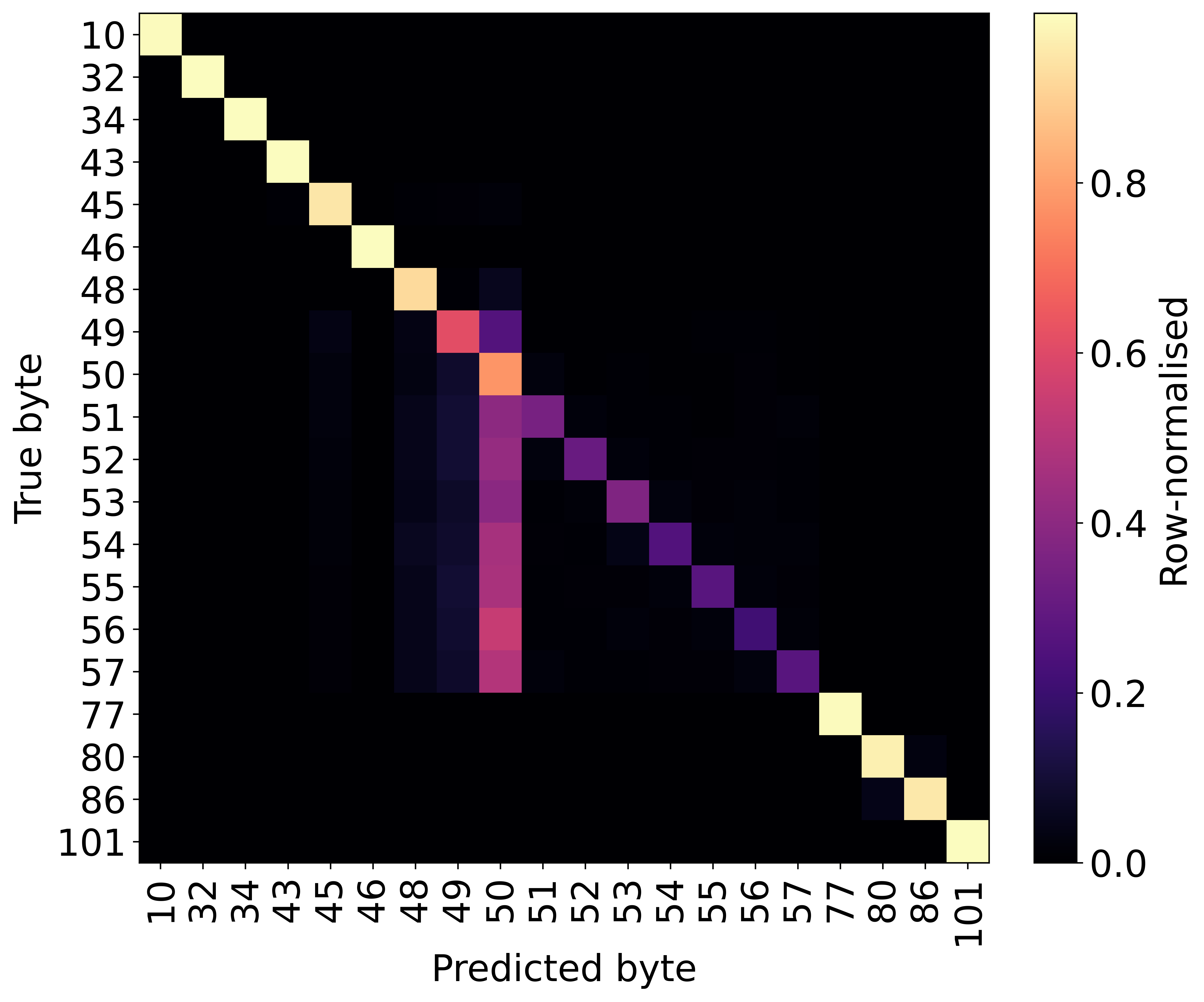}
    \caption{HEPMC}
    \label{fig:ConfMatrix:c}
  \end{subfigure}
  \begin{subfigure}{0.49\linewidth}
    \centering
    \includegraphics[width=0.8\linewidth]{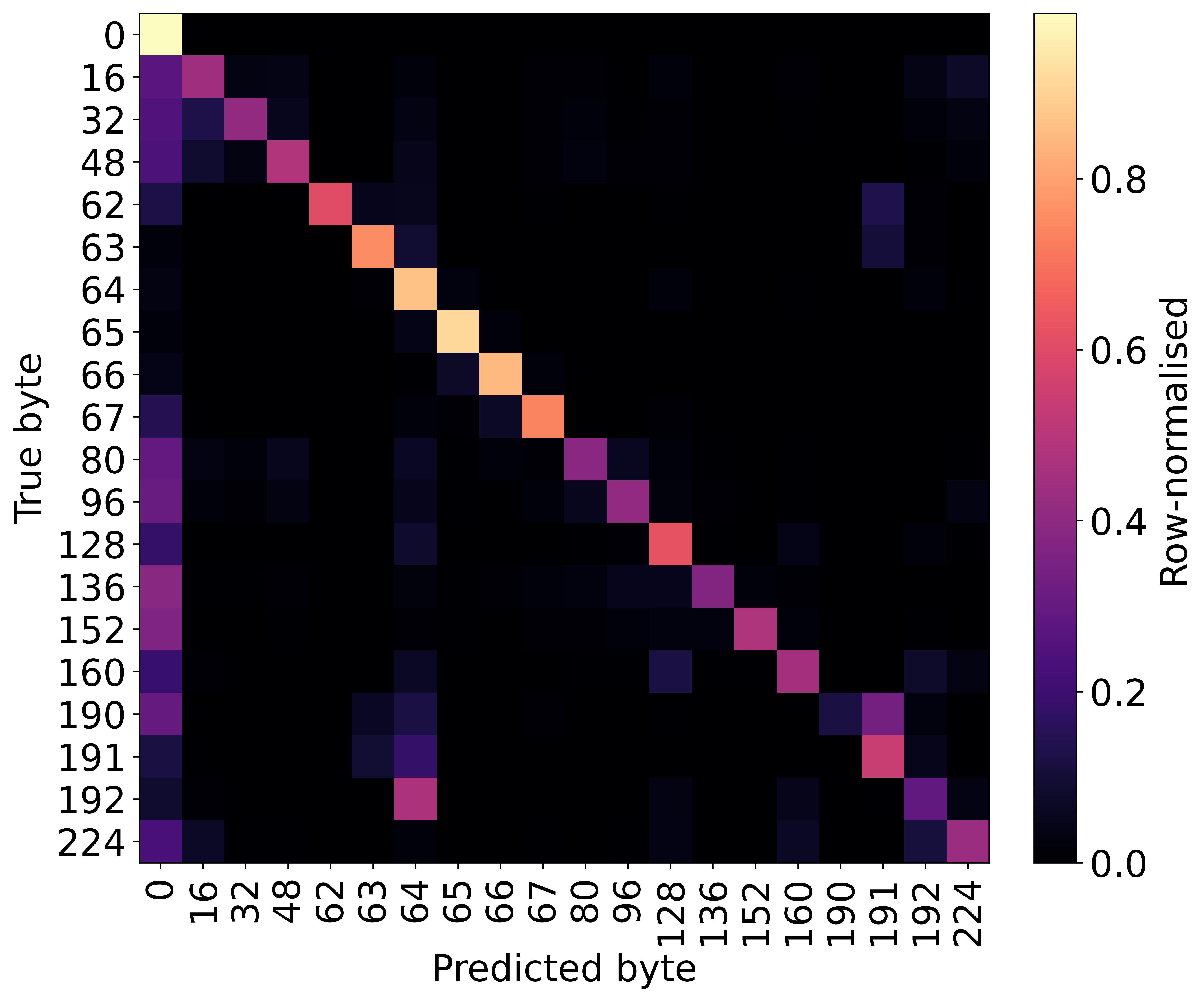}
    \caption{Bundled CMS}
    \label{fig:ConfMatrix:d}
  \end{subfigure}

  \caption{Normalised confusion matrices of the 20 most common bytes: (a) CMS, (b) ATLAS, (c) HEPMC, (d) Bundled CMS.}
  \label{fig:ConfMatrix}
\end{figure*}

\begin{figure*}[ht!]
  \centering
  \begin{subfigure}{0.49\linewidth}
    \centering
    \includegraphics[width=0.8\linewidth]{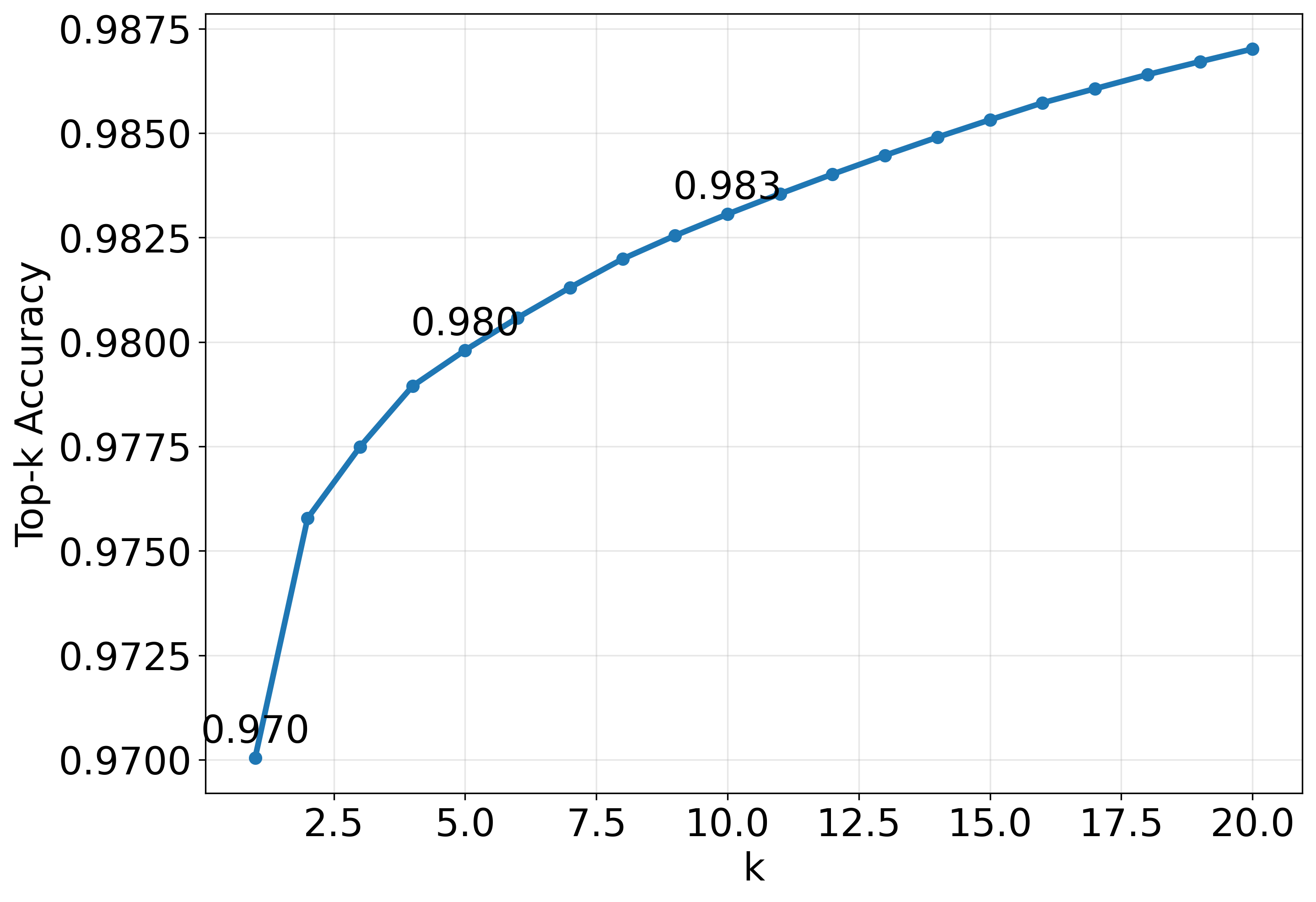}
    \caption{CMS}
    \label{fig:topk:a}
  \end{subfigure}
  \begin{subfigure}{0.49\linewidth}
    \centering
    \includegraphics[width=0.8\linewidth]{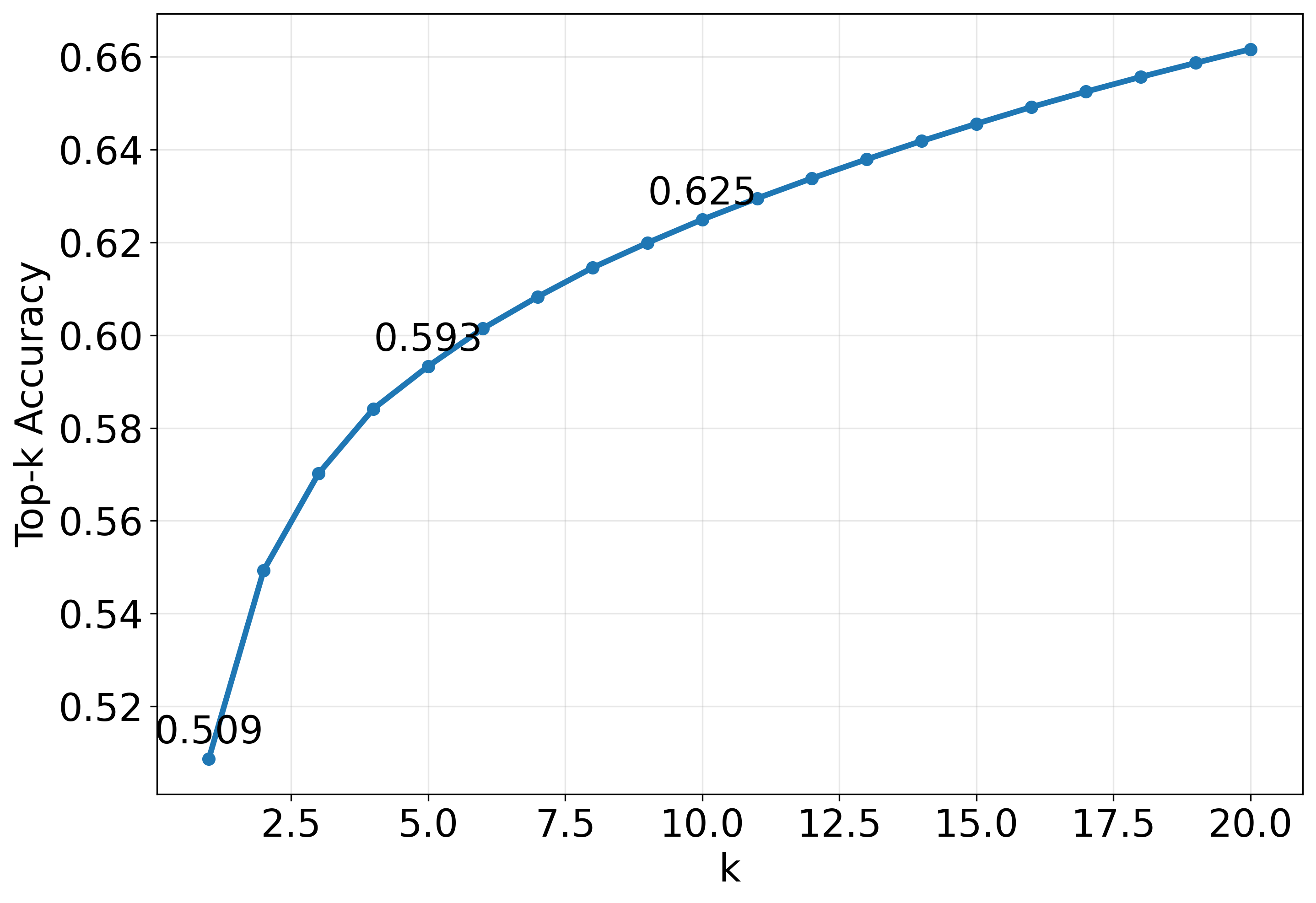}
    \caption{ATLAS}
    \label{fig:topk:b}
  \end{subfigure}

  \vspace{0.6em}

  \begin{subfigure}{0.49\linewidth}
    \centering
    \includegraphics[width=0.8\linewidth]{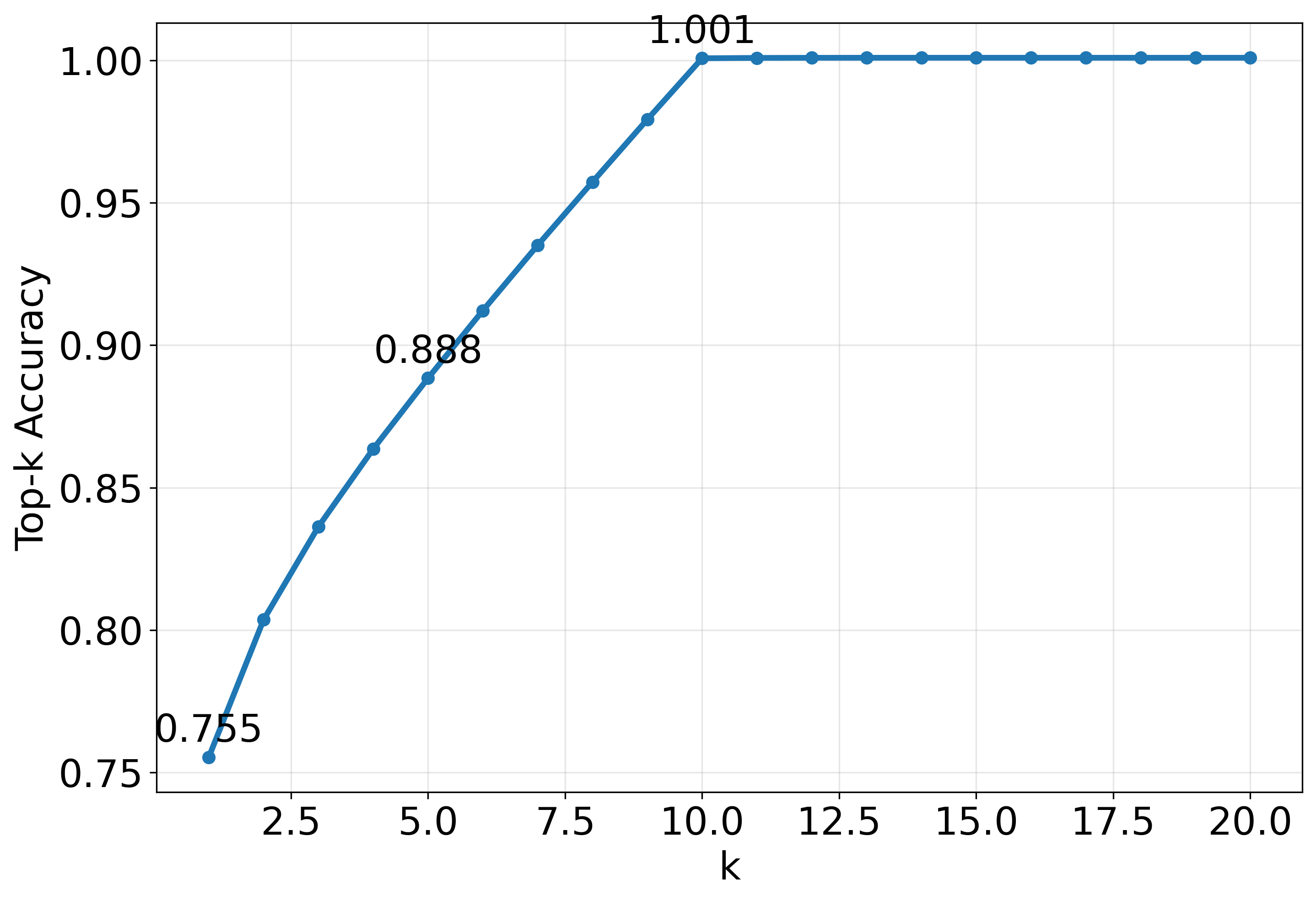}
    \caption{HEPMC}
    \label{fig:topk:c}
  \end{subfigure}
  \begin{subfigure}{0.49\linewidth}
    \centering
    \includegraphics[width=0.8\linewidth]{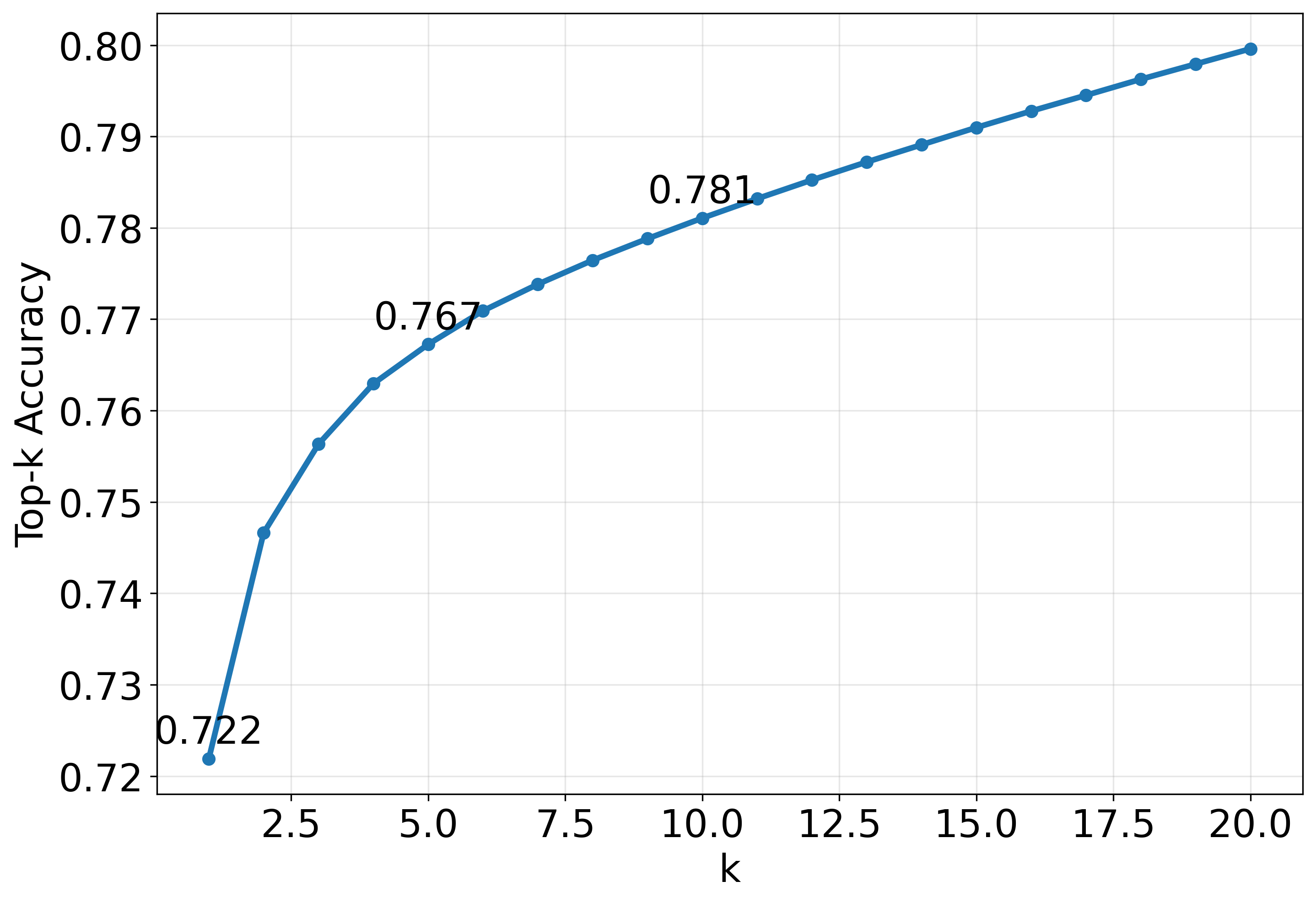}
    \caption{Bundled CMS}
    \label{fig:topk:d}
  \end{subfigure}

  \caption{Top-$k$ prediction accuracy for each dataset: (a) CMS, (b) ATLAS, (c) HEPMC, (d) Bundled CMS. Here, Top-$k$ is the fraction of bytes whose true value lies within the $k$ most probable model predictions; higher Top-1/Top-$k$ imply lower cross-entropy and thus better compression.}
  \label{fig:topk}
\end{figure*}

\begin{figure*}[ht!]
  \centering
  \begin{subfigure}{0.40\linewidth}
    \centering
    \includegraphics[width=\linewidth]{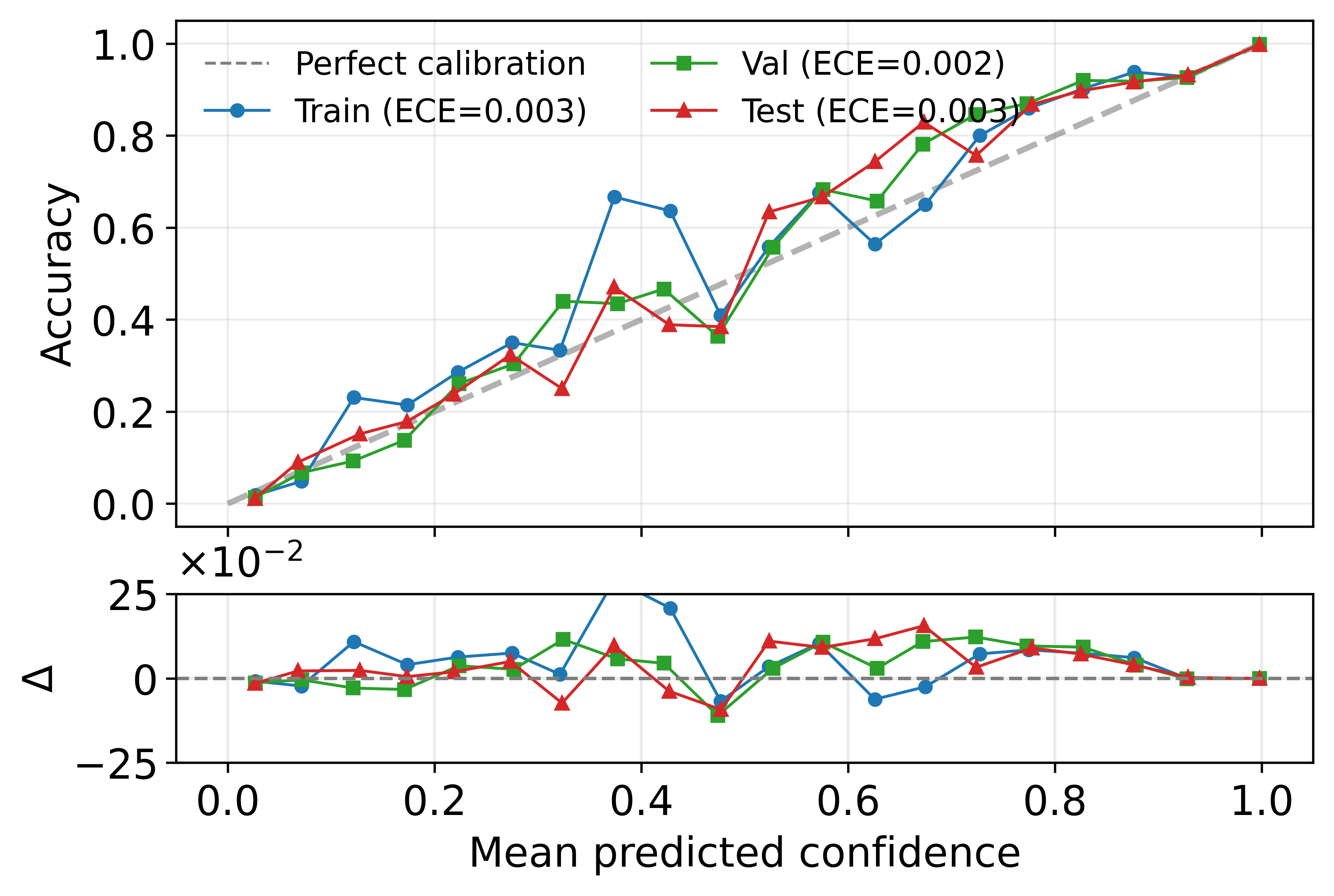}
    \caption{CMS}
    \label{fig:reliability:a}
  \end{subfigure}
  \begin{subfigure}{0.40\linewidth}
    \centering
    \includegraphics[width=\linewidth]{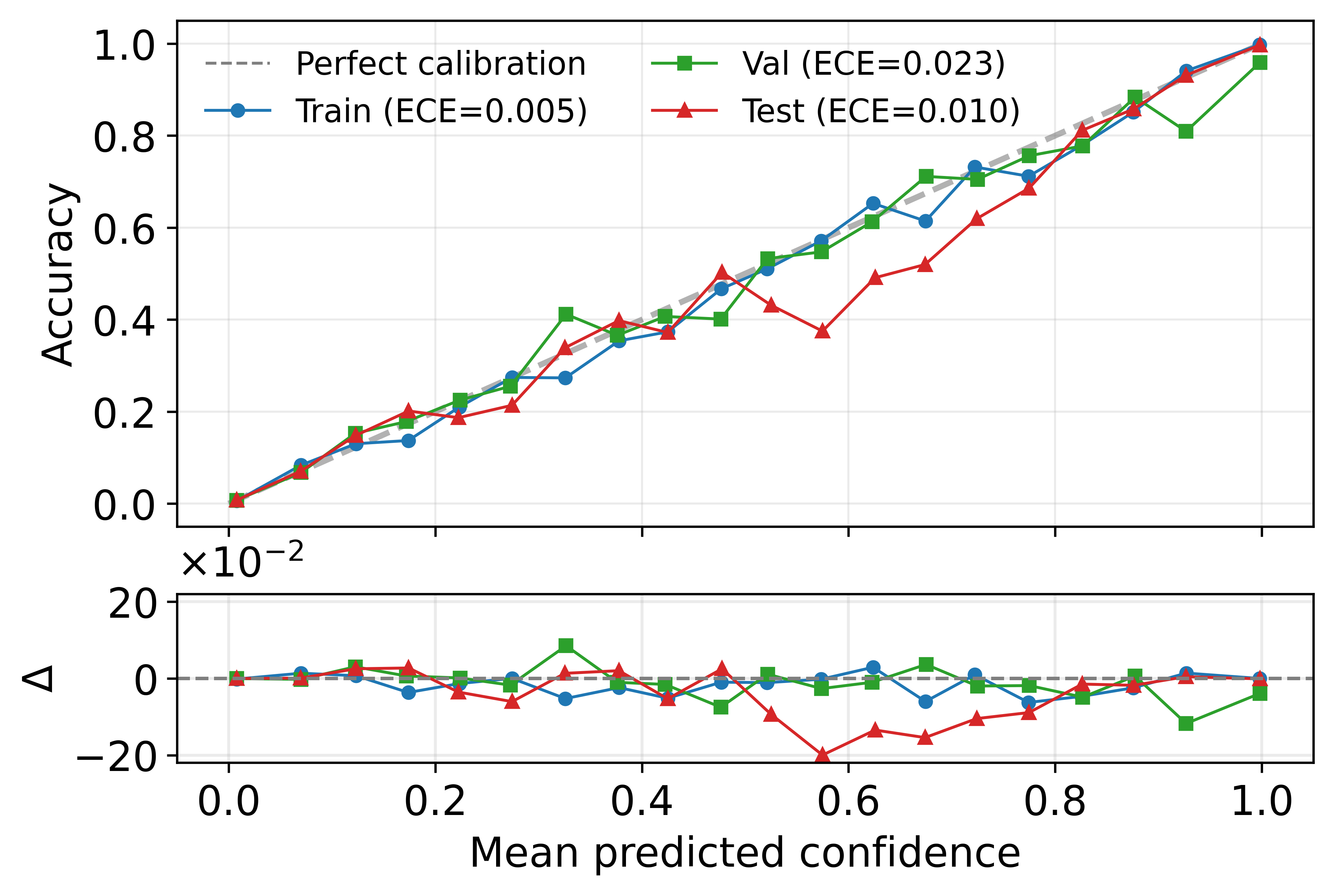}
    \caption{ATLAS}
    \label{fig:reliability:b}
  \end{subfigure}

  \vspace{0.6em}

  \begin{subfigure}{0.40\linewidth}
    \centering
    \includegraphics[width=\linewidth]{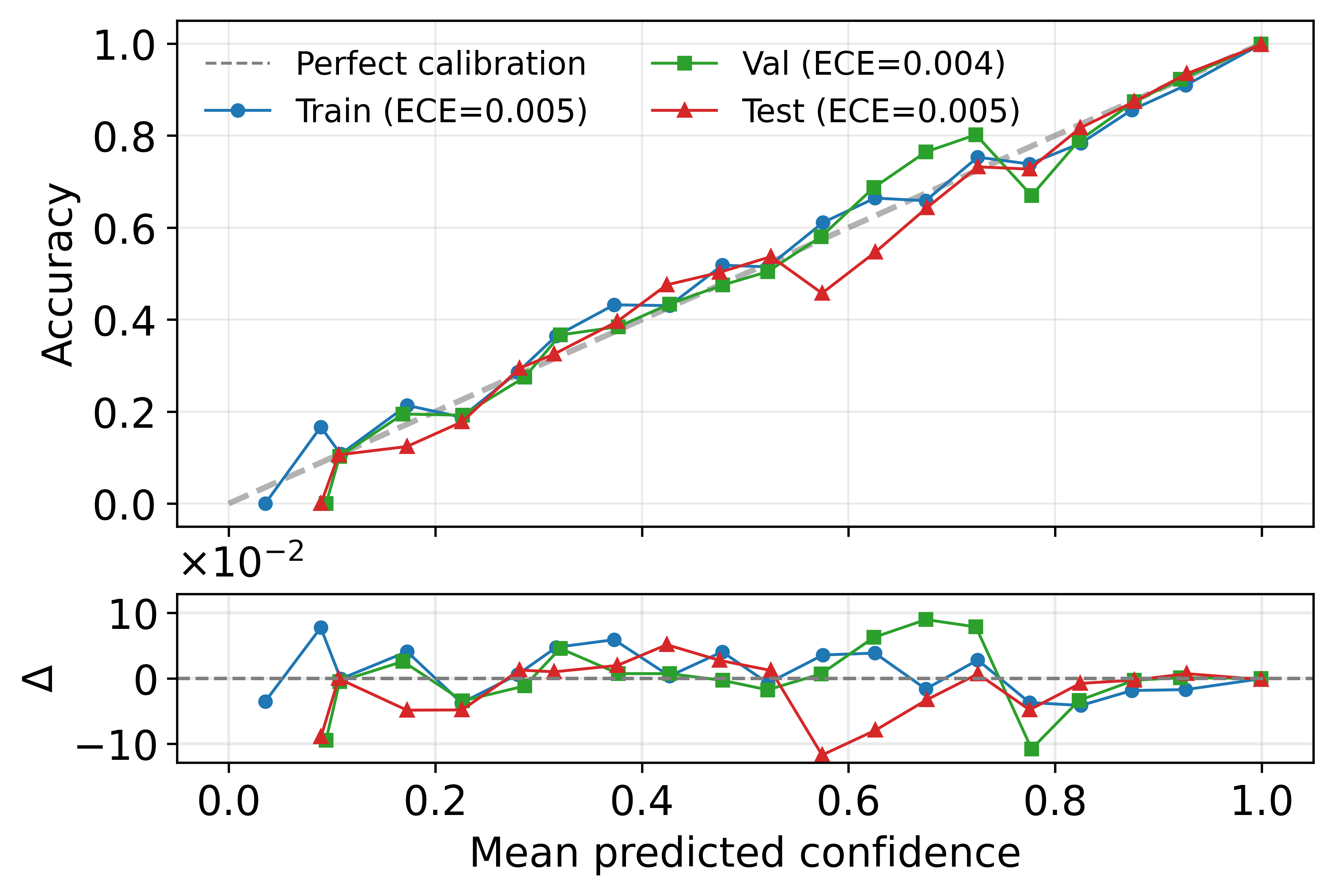}
    \caption{HEPMC}
    \label{fig:reliability:c}
  \end{subfigure}
  \begin{subfigure}{0.40\linewidth}
    \centering
    \includegraphics[width=\linewidth]{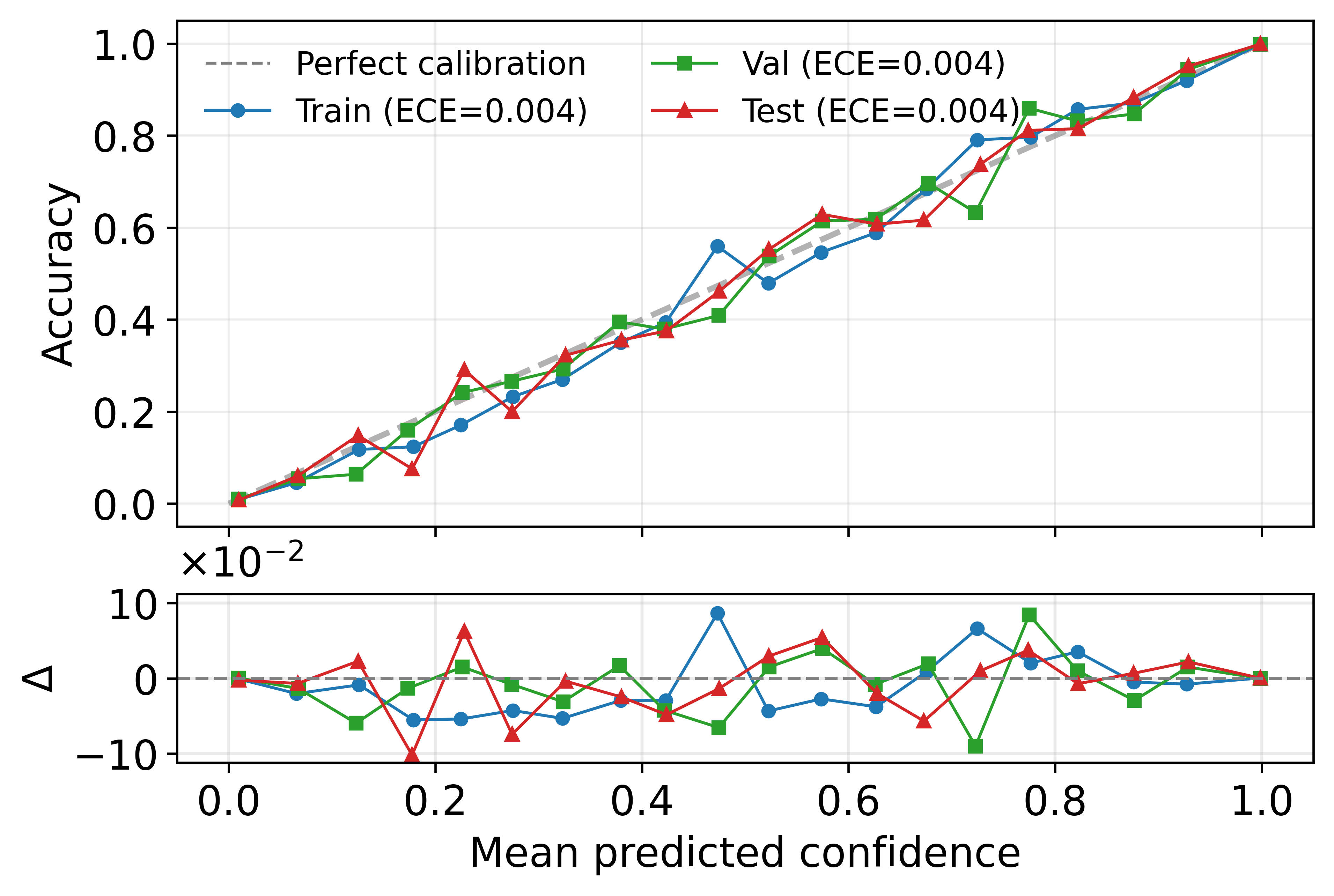}
    \caption{Bundled CMS}
    \label{fig:reliability:d}
  \end{subfigure}

  \caption{Reliability diagrams with residuals for each dataset: (a) CMS, (b) ATLAS, (c) HEPMC, (d) Bundled CMS. $\Delta$ denotes the gap between empirical accuracy and the perfect calibration line.}
  \label{fig:reliability}
\end{figure*}
Confusion matrices, Top-\(k\) curves, and reliability diagrams support the compression outcomes (see Fig.~\ref{fig:ConfMatrix}, Fig.~\ref{fig:topk}, and Fig.~\ref{fig:reliability}, respectively). The Expected Calibration Error (ECE) measures the gap between a model's average predicted confidence and its actual accuracy to see if, for example, predictions made with 80\% confidence are indeed correct 80\% of the time. From a physics perspective, the byte-level confusion matrices show which regions of the feature space the model treats as interchangeable. A sharp diagonal reflects well-resolved structure in the encoded variables, while coherent off-diagonal bands (such as the vertical stripe at byte~0) reveal degeneracies from serialisation, padding, or heavy binning. These patterns directly affect compression, since persistent off-diagonal mass forces $Q$ to spread probability, increasing $D_{\mathrm{KL}}(P\Vert Q)$ and lowering the achievable compression ratio. The same diagnostic can also be used as a data-quality or encoding tool, highlighting poorly encoded columns or value ranges and exposing symmetries or sparsity patterns relevant for downstream analyses.

\paragraph*{a) CMS.}
The confusion matrix is almost perfectly diagonal with a narrow off-diagonal smear. However, it shows a clear vertical band at predicted byte $0$, indicating that many different true bytes (e.g., 8, 16, 24, 48, 96, 128) are frequently misclassified as the zero byte.
Calibration is near-ideal and uniform across splits: ECE$_{\text{train}}=0.003$, ECE$_{\text{val}}=0.002$, ECE$_{\text{test}}=0.003$. Residuals remain within a few percentage points, with slight under-confidence between the 0.3 and 0.9 regions.
Top-$k$ saturates very early, revealing an extremely concentrated posterior: Top-1 $\approx 0.970$, Top-5 $\approx 0.980$, Top-10 $\approx 0.983$, Top-20 $\approx 0.987$. This near-determinism is the driver of the large CMS compression gains: the range coder receives sharply peaked and well-calibrated distributions and therefore codes near the entropy.

\paragraph*{b) ATLAS.}
The top--20 frequent-byte confusion matrix is diagonal-dominant but with a visible vertical band at predicted byte $0$, indicating a prior bias towards the zero token even when the true byte is non-zero. Off-diagonal mass clusters near adjacent values, which suggests local ambiguity within narrow byte families rather than long-range confusions.
Calibration is strong but shows validation drift: ECE$_{\text{train}}=0.005$, ECE$_{\text{val}}=0.023$, ECE$_{\text{test}}=0.010$. Residuals are small and structured: mild under-confidence at low confidence ($\lesssim 0.3$) and notable over-confidence around $0.6$--$0.7$ on test.
Top-$k$ rises slowly, confirming a broader candidate set: Top-1 $\approx 0.509$, Top-5 $\approx 0.593$, Top-10 $\approx 0.625$, Top-20 $\approx 0.66$. These traits explain the moderate compression ratio: probabilities are calibrated but less peaked.

\paragraph*{c) HEPMC.}
The confusion matrix is diagonal with one notable confusion block: true bytes in the $\sim 49$--$57$ range are frequently mispredicted as bytes $49$, $50$, or $51$. Errors are local, within small neighbourhoods, indicating stable token families rather than diffuse mistakes.
Calibration is stable across splits with low error: ECE$_{\text{train}}=0.005$, ECE$_{\text{val}}=0.004$, ECE$_{\text{test}}=0.005$. Residuals oscillate tightly around zero with a small negative dip near $0.6$, implying slight over-confidence at mid-range probabilities.
Top-$k$ climbs quickly then plateaus: Top-1 $\approx 0.755$, Top-5 $\approx 0.888$, and the curve reaches $\approx 1.00$ by $k\approx 10$. Hence the correct byte almost always lies in a compact candidate set, which the coder exploits to achieve a strong ratio.

\paragraph*{d) Bundled CMS.}
The confusion matrix is strongly diagonal dominant, but features a prominent vertical band in the predicted byte $0$, similar to the ATLAS and CMS datasets. This indicates a bias towards predicting the zero byte for many different true byte values (e.g., 16, 32, 48, 80, 96).
Calibration is excellent and highly stable across splits, with very low errors: ECE$_{\text{train}}=0.004$, ECE$_{\text{val}}=0.004$, ECE$_{\text{test}}=0.004$. The residual plot confirms this, showing very small oscillations around the zero line, with no significant systemic over- or under-confidence.
The Top-$k$ curve starts at a strong Top-1 $\approx 0.722$ but rises much more slowly than the original CMS dataset, reaching Top-5 $\approx 0.767$ and Top-10 $\approx 0.781$. This indicates that while the model is often correct, its probability distribution is less concentrated, and the correct byte is often found in a broader set of candidates. This explains why the dataset still compresses well (due to good calibration and a high Top-1) but does not achieve the extreme ratios of the original CMS dataset. However, the core of this difference might lie in the fact that the bundled CMS model was trained with $\sim4\times$ less data than the others. 

Across datasets, small ECE values show that probability calibration is sufficient for efficient entropy coding. Differences in Top-$k$ concentration map directly to the observed compression ratios. 

Targeted mitigations could include context partitioning (e.g.\ per-column models), temperature scaling, or a two-stage head with a sentinel-vs-non-sentinel gate to reduce spurious mass on byte~$0$ without harming calibration.

Relative improvements track byte-level structure after serialisation. CMS exhibits the most exploitable regularity and thus the largest headroom; ATLAS and HEPMC are less structured at the byte level; bundled CMS keeps benefits though float32 reduces redundancy. This matches the cross-entropy framing: reducing \(D_{\mathrm{KL}}(P\|Q)\) lowers the achieved code length under range coding.

\begin{figure*}[ht!]
    \centering
    \includegraphics[width=0.30\linewidth]{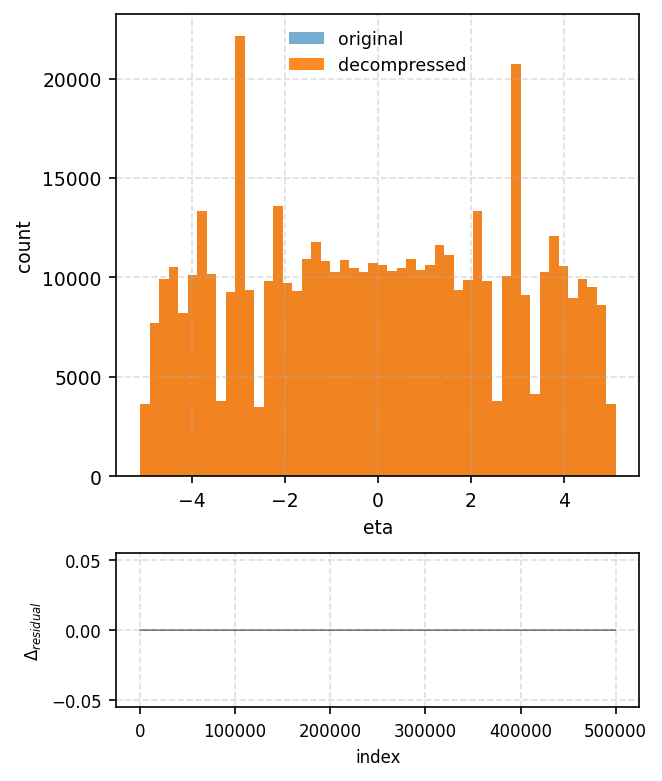}
    \includegraphics[width=0.30\linewidth]{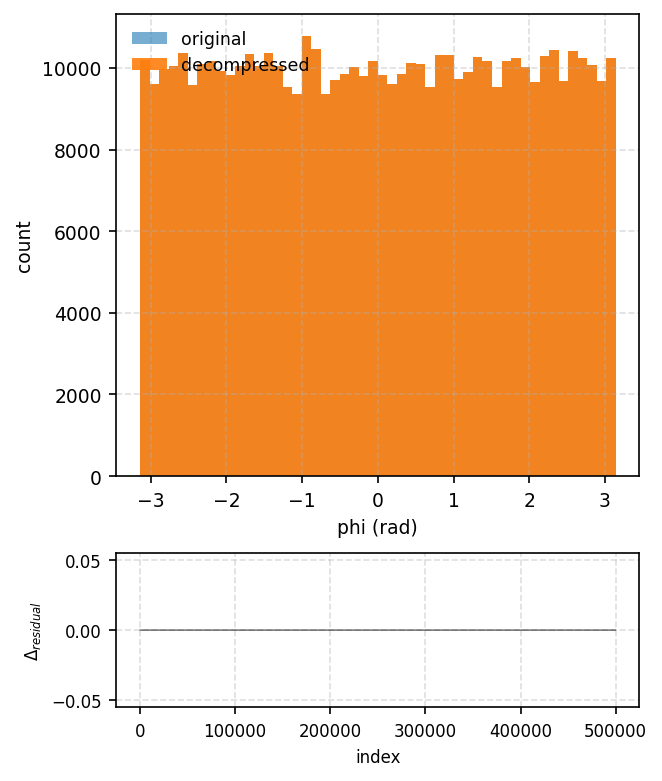}
    \includegraphics[width=0.30\linewidth]{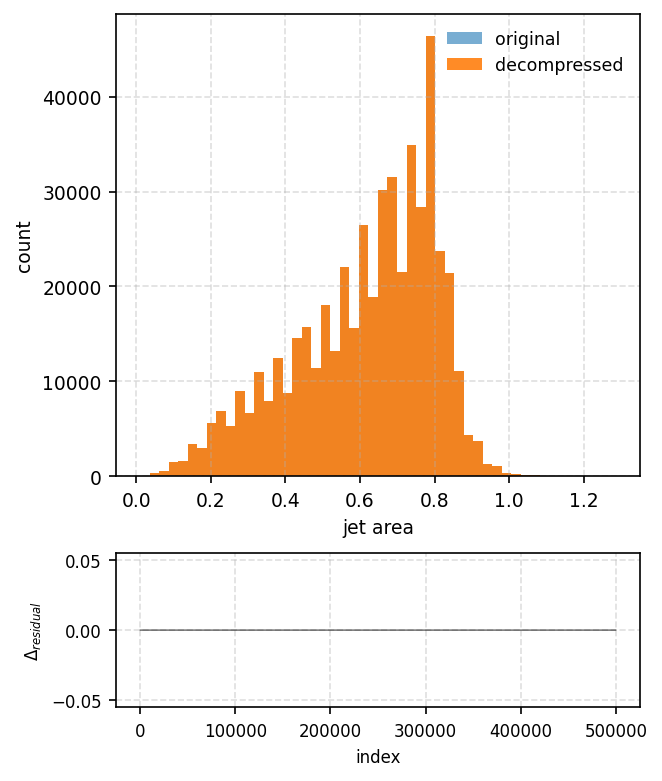}
    \caption{Graphs showing the original and reconstructed columns for a few features ($\eta$, $\phi$ and jet area from left to right) using the Bundled CMS dataset. $\Delta_{\mathrm{residual}} = \text{original} - \text{compressed}=0$ shows that the reconstructed file completely match the original file.}
    \label{fig:origvrecons}
\end{figure*}

Figure~\ref{fig:origvrecons} demonstrates the lossless nature of the compression on the Bundled CMS dataset. It compares the histograms of the original and decompressed data for three distinct physics features: $\eta$ (pseudorapidity), $\phi$ (azimuthal angle), and jet area. In all three plots, the `decompressed' distribution (orange) perfectly overlays and completely obscures the `original' distribution (blue), indicating an identical match. This bit-perfect reconstruction is quantitatively confirmed by the $\Delta_{\mathrm{residual}}$ plots below, which show a flat line at zero, meaning the difference between the original and decompressed values is zero for every entry.

\subsection*{Portability}
Floating-point arithmetic on GPUs makes matrix multiplication (matmul) architecture-dependent, leading to divergences across devices. Consequently, the current PyTorch implementation is not fully portable. It runs on any CPU or CUDA-enabled GPU, but encoding on one machine and decoding on another will require custom kernels and model quantisation. 

Although a number of studies have explored quantisation of Mamba layers \cite{Chiang202558791,xu2025mambaquantquantizingmambafamily}, normalisation and output operations such as \verb|layernorm| and \verb|softmax| are still commonly evaluated in higher-precision floating-point, as is well documented in the Transformer quantisation literature \cite{kim2021ibertintegeronlybertquantization}. These components inherit implementation-dependent IEEE-754 behaviour \cite{osti_976992}, reintroducing non-associativity, kernel-level variation across devices, and hence loss of bitwise determinism in neural compression pipelines. Integer-only schemes such as I-BERT demonstrate that GELU, softmax, and layer normalisation can be implemented in low-precision integer arithmetic for Transformer models \cite{kim2021ibertintegeronlybertquantization}, but an analogous end-to-end deterministic pipeline has not yet been established for Mamba-based compressors. We therefore propose a two-pronged strategy: (i) quantise all Mamba state-space and affine transformations to a bit-deterministic INT8 (or fixed-point) representation with explicitly specified rounding and saturation rules, and (ii) implement custom, deterministic GEMM (general matrix multiplication) kernels and auxiliary routines for the remaining non-quantised components (e.g.\ normalisation and softmax). The use of INT8 arithmetic is expected to increase throughput and reduce memory bandwidth requirements on modern accelerators \cite{Jacob_2018_CVPR}, but any such gains may be partially offset by the overhead and tuning complexity of custom GEMM kernels relative to vendor-optimised libraries \cite{Boehm2022_CUDA_Matmul}. A rigorous characterisation of the net performance impact and determinism guarantees is left to future evaluation.
\section*{Conclusion}
This work has introduced BOA constrictor, a novel, streaming-capable lossless compressor based on the Mamba state space model. We have demonstrated its effectiveness on several High Energy Physics (HEP) datasets, where it achieved state-of-the-art compression ratios, significantly outperforming traditional, widely-used algorithms like LZMA-9 and ZLIB-9. The analysis of the model's predictive behaviour, including its near-perfect calibration and high Top-$k$ accuracy on structured data like the CMS dataset, confirms that the Mamba architecture can effectively learn and exploit deep byte-level regularities, thereby enabling superior compression.

However, despite these highly promising results, we must recognise that the current implementation of BOA is a proof-of-principle and not yet a production-ready algorithm. The primary limitation is practical throughput; while BOA excels in compression ratio, its Storage-Saving Rate ($\sigma_{SSR}$) currently lags behind highly-optimised, CPU-based algorithms like ZLIB. This performance gap highlights that the computational overhead of the neural network, even with GPU acceleration, is a significant bottleneck. Furthermore, as noted in our analysis, the current PyTorch-based implementation faces portability challenges due to its reliance on architecture-dependent floating-point arithmetic for matrix multiplication. This lack of guaranteed bit-level determinism across different hardware is a critical barrier to deployment in a production environment where data must be encoded on one machine and decoded on another, potentially years later.

Consequently, significant future work is required to bridge the gap from promising prototype to deployable tool. The validation presented here, while robust, has been limited to a few specific, pre-selected datasets. More extensive testing is needed across a wider variety of HEP data formats, as well as data from other scientific domains, to fully assess the algorithm's generality.

Most critically, future work must prioritise comprehensive tests in a real streaming environment. This would involve integrating BOA into a simulated or operational data acquisition (DAQ) or analysis framework to measure its performance against live data sources. Only then can we truly evaluate its real-world throughput, latency, and resilience under the demanding, high-velocity conditions of HEP experiments.

In conclusion, while BOA is not an ``out-of-the-box" solution, it provides compelling evidence that Mamba-based models represent a solid and powerful new approach to lossless compression. The path forward is clear: future development must focus on rigorous performance optimisation, potentially through custom kernels and model quantisation to address both speed and portability. Further investigation into energy-per-terabyte processed will also be a critical metric for a field where energy costs and environmental impact are a major concern. If these engineering challenges are properly handled, this methodology holds significant promise for setting a new standard in scientific data compression and helping to manage the immense data volumes of the future.
\acknowledgments
This work was funded by EPSRC Studentship EP/W524347/1 (Project 2932638) via MADSIM CDT, the University of Manchester Dame Kathleen Ollerenshaw Fellowship, and is part of a project that has received funding from the European Research Council under the European Union’s Horizon 2020 research and innovation program (grant agreement 101002463).
\textbf{All} analysis was done on the same computer with an Nvidia\texttrademark{} RTX 5090, an Intel\texttrademark{}  Core 7 265k, and 192GB RAM. However, portability tests were carried out on another computer with an Nvidia\texttrademark{}  RTX 3060, Intel\texttrademark{}  i7-11370H, and 8GB RAM.

The code for this algorithm including the bundled example is available on Zenodo \cite{akshatgupta}.

\bibliographystyle{apsrev4-2}

\bibliography{ref}

\clearpage
\appendix
\appsection{Mamba Background}{mambabg}
SSMs are a class of models that map a 1D input sequence $x(t)$ to an output $y(t)$ via a latent state vector $h(t) \in \mathbb{R}^N$. The underlying system is described by a set of linear Ordinary Differential Equations (ODEs) \cite{gu2022efficientlymodelinglongsequences}:
\begin{align}
    \dot{h}(t) = \mathbf{A}h(t) + \mathbf{B}x(t),\\
    y(t) = \mathbf{C}h(t) + \mathbf{D}x(t),
\end{align}
where, $A$ is the state matrix, $B$ is the input matrix, $C$ is the output matrix, and $D$ is the feed-through matrix. The state $h(t)$ acts as a compressed representation of the history of the input $x_{<t}$.

This should be discretised for digital computation. Hence, given a sampling time-step $\Delta$, we convert the continuous parameters ($A$,$B$) to discrete parameters ($\bar{A}$, $\bar{B}$). This Linear Time-Invariant (LTI) SSM uses the Zero-Order Hold (ZOH) model, giving \cite{gu2022efficientlymodelinglongsequences}:
\begin{align}
  \bar{\mathbf{A}} = \exp(\Delta \mathbf{A}), \\
\bar{\mathbf{B}} = (\exp(\Delta \mathbf{A}) - \mathbf{I})\Delta\mathbf{A}^{-1}\Delta\mathbf{B},
\end{align}
leading to a discrete-time linear recurrence relation:
\begin{align}
h_k = \bar{\mathbf{A}}h_{k-1} + \bar{\mathbf{B}}x_k \label{discretelin1}\\
y_k = \mathbf{C}h_k + \mathbf{D}x_k. \label{discretelin2}
\end{align}
This recurrent formulation is highly efficient for inference, requiring $O(L)$ time and $O(1)$ memory (beyond model parameters) to process a sequence of length $L$ (see Appendix \ref{appendix:proof01}). 

However, these dynamics are fixed for all inputs. The Mamba architecture overcomes this by introducing a selection mechanism that allows the model's parameters to be conditioned on the input data \cite{gu2024mambalineartimesequencemodeling}. Instead of using fixed matrices, it makes the timestep and the matrices $B$ and $C$ functions of the current input $x_t$. This is achieved by passing the input through linear projection layers:
\begin{align}
    \Delta_t = \text{softplus}(W_{\Delta}x_t + b_{\Delta}), \\
 B_t = W_{B}x_t + b_{B},\\
 C_t = W_{C}x_t + b_{C}, 
\end{align}
where, $\text{softplus}$ ensures that $\Delta_t$ is always positive, and $W$ and $b$ are the weight matrices and bias vectors respectively. Here, a small $\Delta_t$ means that the system changes slowly, which implies that the previous context is important and vice versa. These input-dependent parameters ensure that Mamba can selectively remember or forget information from the history based on the current context. 

This input dependence breaks the time invariance required for a simple convolutional representation during training. This is rectified by employing a parallel scan algorithm, which allows the recurrent computation to be executed efficiently on modern hardware.\\
\appsection{Proof of Complexity for LTI-SSM}{appendix:proof01}
Let $C_T(L)$ be the total computational cost for a sequence of length $L$. This cost is the sum of the costs incurred at each step $k$:
\begin{align}
    C_T(L) = \sum_{k=1}^L C_k,
\end{align}
where, $C_k$ is the cost of computation at step $k$. The operations at each step consist of matrix-vector products and additions. In Eq. \ref{discretelin1}, the cost is dominated by $O(N^2)$ while in Eq. \ref{discretelin2}, it is dominated by $O(EN+EF)$ given that $\bar{A} \in \mathbf{R}^{N \times N}$, $\bar{B} \in \mathbf{R}^{N\times F}$, $\bar{C} \in \mathbf{R}^{E\times N}$, and $\bar{D} \in \mathbf{E}^{N\times F}$. However, $N, E,$ and $F$ are constant with respect to the sequence length L. Therefore, the cost per step $C_k$ is a constant, and hence, $C_T = L.C_k = O(L)$ time complexity.

A similar argument is made for the memory complexity where the auxiliary memory $M_{\mathrm{aux}}$ required for the computation is given by $O(N+E+F)$ at any step. However, since they are again fixed, this boils down to $O(1)$ memory complexity.

\appsection{File Subset Details}{whybaler}
We convert all dataset to binary file so that we do not need to write a special loading algorithm for individual experiments. The bundled CMS Dataset is taken from the Baler Repository \cite{bengtsson2024balermachinelearning}. Here, we keep only the \verb+data+ array so that we can store the data as binary more easily. Furthermore, we reduced the data format to \verb+float32+ to halve the size for easier training and easier upload to GitHub. We provide conversion files from and to the original formats for all the other datasets.


\end{document}